\documentclass[citc]{PoS}

\usepackage{amsmath}
\usepackage{multirow}
\newcommand{\Red}[1]{{\bf {#1}}}

\title{Status of 2+1 flavor, $32^3\times 64$ domain wall fermion simulations}

\ShortTitle{Status of 2+1 flavor, $32^3\times 64$ domain wall fermion simulations}

\author{\speaker{Chulwoo Jung }
        for RBC and UKQCD collaborations\\
		Physics department, Brookhaven National Laboratory, Upton, NY 11973, U.S.A.\\
        E-mail: \email{chulwoo@bnl.gov}}

\abstract{We describe details of $ a ^{-1} \sim 2.2$Gev, $L \sim 3fm$ dynamical domain wall fermion simulations which will allow us to do a more systematic continuum 
extrapolation in combination with existing simulations. 
Details of the simulations such as algorithm choices and machine performance, 
as well as results of basic measurements are presented. These configurations 
are presently being generated on the QCDOC machine at Edinburgh and the DOE 
QCDOC machine at Brookhaven as part of a joint project with LHPC.}

\FullConference{The XXV International Symposium on Lattice Field Theory\\
		 July 30-4 August 2007\\
		 Regensburg, Germany}

\begin{document}

\section{Introduction}
Continuing theoretical advances in lattice gauge theory, especially in chiral fermion formulations and fermion simulation algorithms, and increasing computational resources are making systematic continuum extrapolation 
of many QCD quantities without uncontrolled systematic error a reality. 
RBC and UKQCD collaborations have generated dynamical 2+1 flavor Domain Wall Fermion (DWF) ensembles with $a^{-1}\sim 1.7$Gev \cite{DWF16Ls16,DWF16Ls8,DWF24}, which has allowed extrapolation in quark mass and lattice volume. 
Table~\ref{table:evol} is a list of existing 2+1 flavor DWF ensembles.

Gauge ensembles with a smaller spacing is the obvious next step in making 
continuum extrapolations more systematic. 
To this end, RBC and UKQCD collaborations started generating $\beta= 2.25, 32^3 \times 64 \times 16$ dynamical DWF configurations with 2 different light quark masses. 
Recently, LHPC collaboration joined this effort and now part of the ensembles are being generated with the joint allocation on DOE QCDOC at Brookhaven Notional Laboratory as a result.
We are aiming at $a^{-1}\sim 2.2$Gev, $m_{PS} L >4 $, which will allow us to get the statistical and systematic errors down to a few percent level
for the lattice studies of quantities such as weak matrix elements and hadron matrix elements.
\begin{table}
\begin{center}
	\begin{tabular}{c|c|c|c|c|c|c|c}
		\hline
  		$\beta$ & $L/a$ & $m_sa$ & $m_la$ & $\hat m_s/\hat m_l$  & $m_{PS} L $ & $\tau$(MD) & Accept.\\
		\hline
		\multirow{7}{*}{2.13} & 
		\multirow{3}{*}{$16^3 \times 32 \times 16$} & 
		\multirow{3}{*}{0.04} & 0.01 & 3.3 & 3.9 & 4000 & \Red{57\%} \\
		&& & 0.02  & 1.86 &  5.2 & 4000  & \Red{56\%}\\
		&& & 0.03  & 1.3  &  6.2 & 7500  & 82\% \\
		\cline{2-8}
		&\multirow{4}{*}{$24^3 \times 32 \times 16$} & 
		\multirow{4}{*}{0.04} & 0.005 & 5.4 & 4.6 & 6518+ & 73\%\\
		&& & 0.01  & 3.3 & 5.9  & 4700 & 70\% \\
		&& & 0.02  & 1.86 & 7.8 & 2800 & 71\% \\
		&& & 0.03  & 1.3  & 9.3 & 2800 & 72\% \\
		\hline
		\multirow{2}{*}{2.25} & 
		\multirow{2}{*}{$32^3 \times 64 \times 16$} & 
		\multirow{2}{*}{0.03} & 0.004 &$\sim 6.6$ & $\sim 4.1$ & 1628+ & 72\% \\
		 &&  & 0.006 & $\sim 4.6$&   & 1508+ & 75\% \\
		\hline
	\end{tabular}
\end{center}
\caption{ (2+1) flavor dynamical DWF ensembles generated by RBC and UKQCD collaborations. $\hat m_{\{l,s\}} = m_{\{l,s\}}+m_{res}$.
The first 2 ensembles with acceptance in boldface are generated with a different variant of Rational hybrid Monte Carlo (RHMC\cite{RHMC}) (RHMC I in \cite{DWF16Ls16}). $\tau$(MD) denotes the total trajectory length in MD units and the numbers with "+" denotes ongoing productions.}
\label{table:evol}
\end{table}

A detailed description of simulation algorithm and performance is given in section~\ref{section:sim} and basic quantities and preliminary mass measurements on $m_{l}=0.004$ ensemble are presented in section~\ref{section:meas}. 

\section{Simulation details}
\label{section:sim}
As described in \cite{DWF16Ls16,DWF16Ls8,DWF24,local}, we use  the combination of the DWF formulation from Furman and Shamir \cite{shamir} and
Iwasaki gauge action, which is shown to suppress lattice dislocations enough to give DWF good chiral symmetry while allowing for enough topology tunneling for the range of lattice spacings we are interested in.


The simulation of 2 light and 1 strange quarks is actually done as a 
combination of 
(1+1+1) flavor of strange quarks, done with rational quotient approximation, and 2 flavors of light quark preconditioned by the strange quark\cite{hasenbusch}. 
 While the preconditioning mass does not have to be the same as the strange quark, we found the strange quark is close to be optimal as the preconditioning mass in DWF simulations on smaller volumes.
Using $ {\cal D}(m_f) =  D^\dagger_{DWF}(M_5,m_f) D_{DWF}(M_5,m_f)$ where
$M_5$ is the domain wall height, fixed at 1.8, and $m_f$ is the DWF mass term,
the fermion determinant with the corresponding Pauli-Villars fields 
can be written as
{
\begin{align*}
\int \left[ dU \right] \mbox{exp}\left(-\left(S_F[U]+S_{PV}[U]\right)\right) &= \mbox{det} 
 \left[ \frac{ {\cal D}(m_s)^{1/2}{\cal D}(m_l)} {{\cal D}(1)^{3/2}} \right] 
=  \mbox{det} \left[\frac{ {\cal D}(m_s)} {{\cal D}(1)} \right]^{3/2}
  \mbox{det} \left[\frac{ {\cal D}(m_l)} {{\cal D}(m_s)} \right] \\
\sim    
\mbox{det} \left[ {\cal R}_{\frac12} \left(\frac{ {\cal D}(m_s)} {{\cal D}(1)}\right) \right] &
\mbox{det} \left[ {\cal R}_{\frac12}  \left( \frac{  {\cal D}(m_s)} {{\cal D}(1)}\right)\right] 
\mbox{det} \left[ {\cal R}_{\frac12} \left(\frac{ {\cal D}(m_s)} {{\cal D}(1)}\right) \right]
 \mbox{det} \left[\frac{ {\cal D}(m_l)} {{\cal D}(m_s)} \right]
\end{align*}
}
Where ${\cal R}_{a}(x)$ denotes rational approximation of $x^a$ and each determinant term is evaluated by separate pseudofermions.
Omelyan integrator\cite{omelyan} with $\lambda=0.22$ is used in each level of multiscale integrators with $N_{step} = 16$, 
 $\Delta t_{light}  : \Delta t_{heavy}  : \Delta t_{gauge}  = 1/8: 1/8: 1/48$.

 The suppression of force from light quarks from Hasenbusch preconditioning allows us to have the light quark have the biggest step among different terms (the nature of higher-order integrator such as Omelyan effectively makes $\Delta t_{heavy}$ half of $\Delta t_{light}$), decreasing the computational cost significantly. Also, we decided to simulate with trajectory length $\tau= 2$  to make configurations possibly decorrelate more effectively. 

The combination of higher-order, multiscale  integrators and (rational) quotient terms makes the evolution program a  heavily nested one. One way to describe this is
{\large
\begin{align*}
\tau = 2 = 6 MInv + 1 CG + [ [ 12GF+ [ 3 MInv & + 2RF ] \times 3 ] \times 2 + 12GF + 1 CG + HF ] \nonumber \\
 \times 32  +[12GF+ [3MInv  + 2RF] & \times  3]\times 2 + 12GF  + 6 MInv + 1CG
\end{align*}
}
Where $MInv, CG, GF, RF,$ and $ HF$ denote multimass solver for rational quotient terms, inverter for preconditioned light quarks, gauge force, 
rational quotient fermion force and quotient fermion force respectively.
 Expanding the expression without changing the order or terms gives all the computational routine in an MD trajectory in order.


Algorithm described above is implemented and fully optimized for QCDOC in Columbia Physics System(CPS\cite{cps}). 
All of the production runs are done on 4096-node QCDOC partitions at Brookhaven National Laboratory and another 4K partition at Edinburgh Parallel Computing Center. Each partitions are running at 400MHz, which gives 800MFlops/s peak per processor.

Table~\ref{table:perf} shows performances of each routines in the $24^3$ and $32^3$ DWF evolution.
The multimass solver for rational quotient part of the action ($MInv$) is the 
dominating part, especially for relatively heavy light quarks $(m_s/m_l <4)$. 
While the large number of nodes in each partition and a feature of CPS which allows only even number of sites on each nodes makes it necessary to split the 5th dimension and make strictly 4 dimensional routines such as gauge force duplicate calculation along the 5th dimension in some cases, the effect is at the level of a percent of the total time.

The sum of time on individual routines are slightly less than the total time ($\sim$ 5\% of the total time for $32^3$ ensembles). The most of the descrepancy is from the eigenvalue measurement routines which are run at the time of each Metropolis step to check if the eigenvalues of DWF dirac operators are witin range of the rational quotient approximation. 
While the performance of the routine is expected to be 
close to that of inverters, it was not measured and we did not include the flops for the numbers included in the table. 
As a result, the overall performance slightly less than 200MFlops/s per processor, 25\% of the peak.
A more detailed analysis of mass scaling of each routines can be found in \cite{nhc}. 
\begin{table}
\hspace{-1.0cm}
\begin{center}
\begin{tabular}{c|c|c|c|c|c|c}
\hline 
\multicolumn{7}{c} { $24^3\times 64 \times 16(m_s=0.04)$, 
Local volume = $6^3\times 2 \times 8$}  \\
\hline 
& $m_l=$0.03 & 0.02  & &0.01 &0.005 &\\
\hline 
Routines 
& time(s) & time(s) & MFlops/s &
time(s)  & time(s) & MFlops/s\\
\hline
MInv
 & 1225 & 1213 & 221 & 1195 & 1367 & 225 \\
CG
 & 173 & 223  & 273 & 370 & 634  & 258 \\
GF
 & \Red{60}  & \Red{60}   & 257 & \Red{62}  & \Red{73} & 250 \\
RF
 & 218 & 218  & 36  & 232 & 274 & 34 \\
HF
 & 10  & 10   & 4.5 & 10  & 12 & 4.5 \\
\hline
Total time(seconds) & 1941 & 1983 &  & 2124 & 2635  &   \\
\hline 
Total MFlops/core& 333557 & 344555 &  & 380018 & 489642  &   \\
\hline 
Total flops($\times 10^{12}$)& 1366 & 1411 &  & 1557 & 2006  &   \\
\hline 
\end{tabular}
\begin{tabular}{c|c|c|c|c}
\multicolumn{5}{c} { } \\
\hline 
\multicolumn{5}{c} { $32^3\times 64 \times 16 (m_s=0.03)$ } \\
\hline 
$m_l$& \multicolumn{2}{c} { 0.006 } & \multicolumn{2}{|c} { 0.004} \\
\hline
Local volume& \multicolumn{2}{c} { $8^3\times 2 \times 8$ } & \multicolumn{2}{|c} { $4^3\times 8 \times 16$ } \\
\hline
Routines & time(sec) & MFlops/s & time(sec) & MFlops/s \\
\hline
MInv &  5062  & 172 & 4263 & 205 \\
CG            &  1964  & 213 & 2038 & 268 \\
GF   &  \Red{214}   & 256 & 104  & 263 \\
RF    &  1130  & 25  & 939  & 28  \\
HF     &  39    & 4.3 & 10   & 16 \\
\hline
Total time(seconds)       &  9035 &  & 7733 &  \\
\hline 
Total MFlops/core & 1344806 &  & 1473903  &\\
\hline 
Total flops ($\times 10^{12}$) & 5467 &  & 6037  &\\
\hline
\end{tabular}
\end{center}
\caption{Performance of computation routines in DWF RHMC on QCDOC. Bold numbers 
denotes 4-dimensional routines which are duplicated along 5th dimension when the 5th dimension is split.}
\label{table:perf}
\end{table}

\section{Basic measurements}
\label{section:meas}
\begin{figure}
\begin{minipage}{0.5\textwidth}
\begin{center}
\includegraphics[angle=270,width=3.3in]{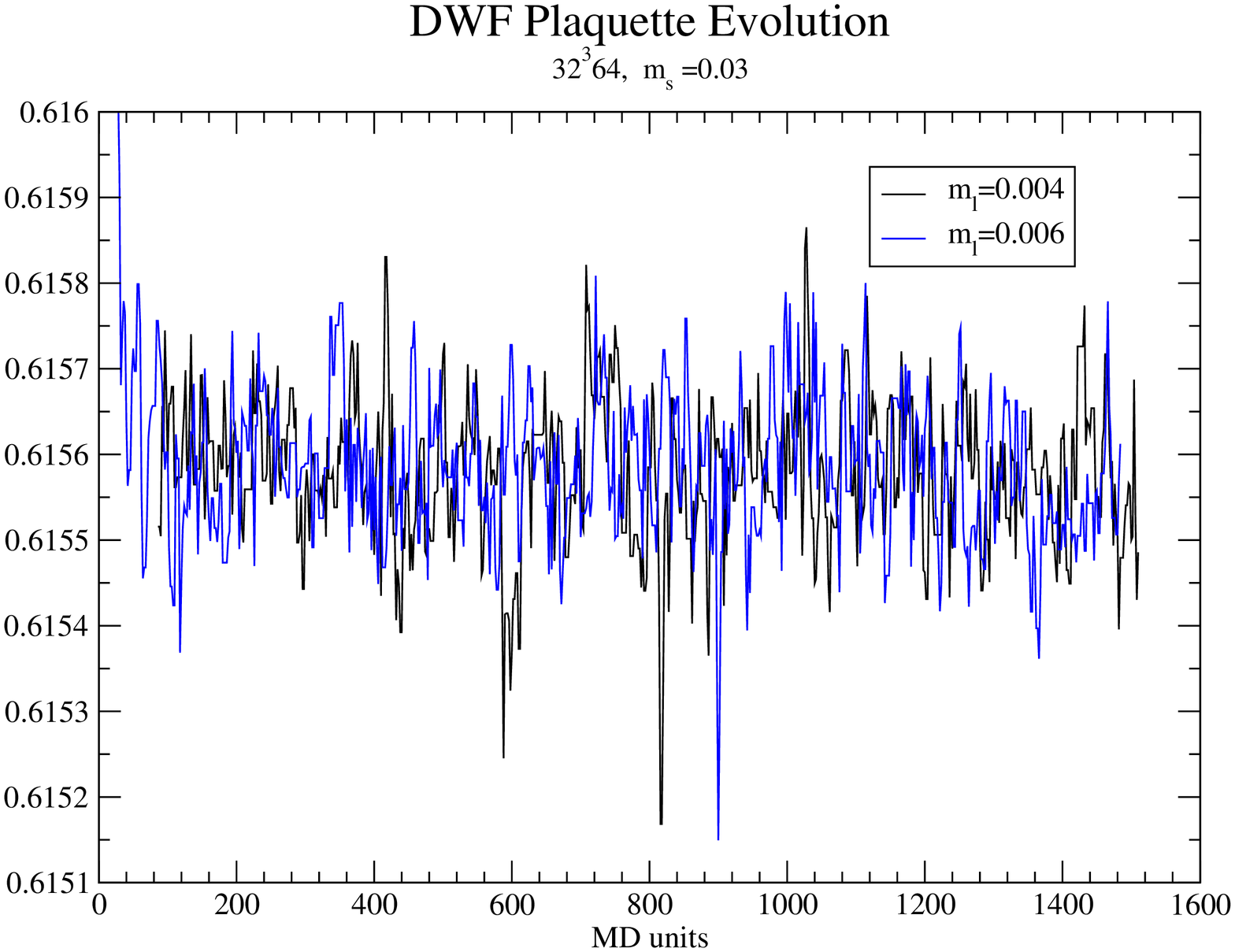}
\end{center}
\end{minipage}
\begin{minipage}{0.5\textwidth}
\begin{center}
\includegraphics[angle=270,width=3.3in]{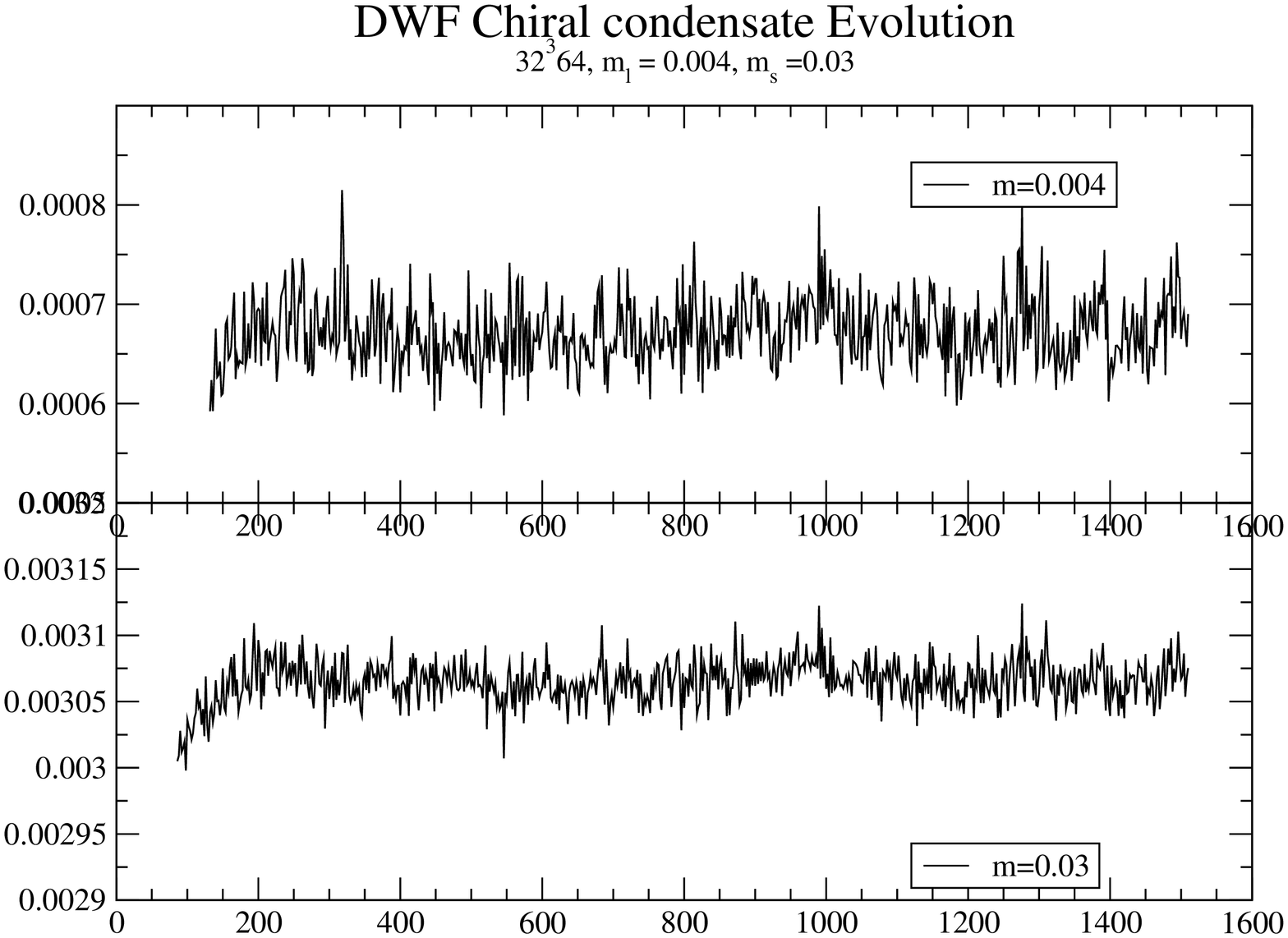}
\end{center}
\end{minipage}
\caption{Evolution of the average plaquette and the chiral condensate for
$\beta=2.25, 32^3\times64\times 16$ ensembles.
Average plaquette $\langle P \rangle (m_l=0.004)=0.615574(13)$ and
$\langle P \rangle (m_l=0.006)=0.615591(9)$. 
$<\bar\psi \psi >$ shown here is from 
$m_l=0.004$ ensemble.} 
\label{fig:evol}
\end{figure}

Figure~\ref{fig:evol} shows the evolution of the plaquette and the chiral condensate. The time series analysis of these quantities show they have the autocorrelation time of 7-14 MD units, which is smaller than what is reported in \cite{DWF16Ls16} from meson correlators.
Measurements of meson correlators over more configurations than what is available is needed to compare how effectively the RHMC algorithm is generating decorrelated lattice configurations.
\label{section:evol}

\begin{figure}
\begin{minipage}{0.5\textwidth}
\begin{center}
\vspace{-5mm}
\includegraphics[angle=270,width=3.3in]{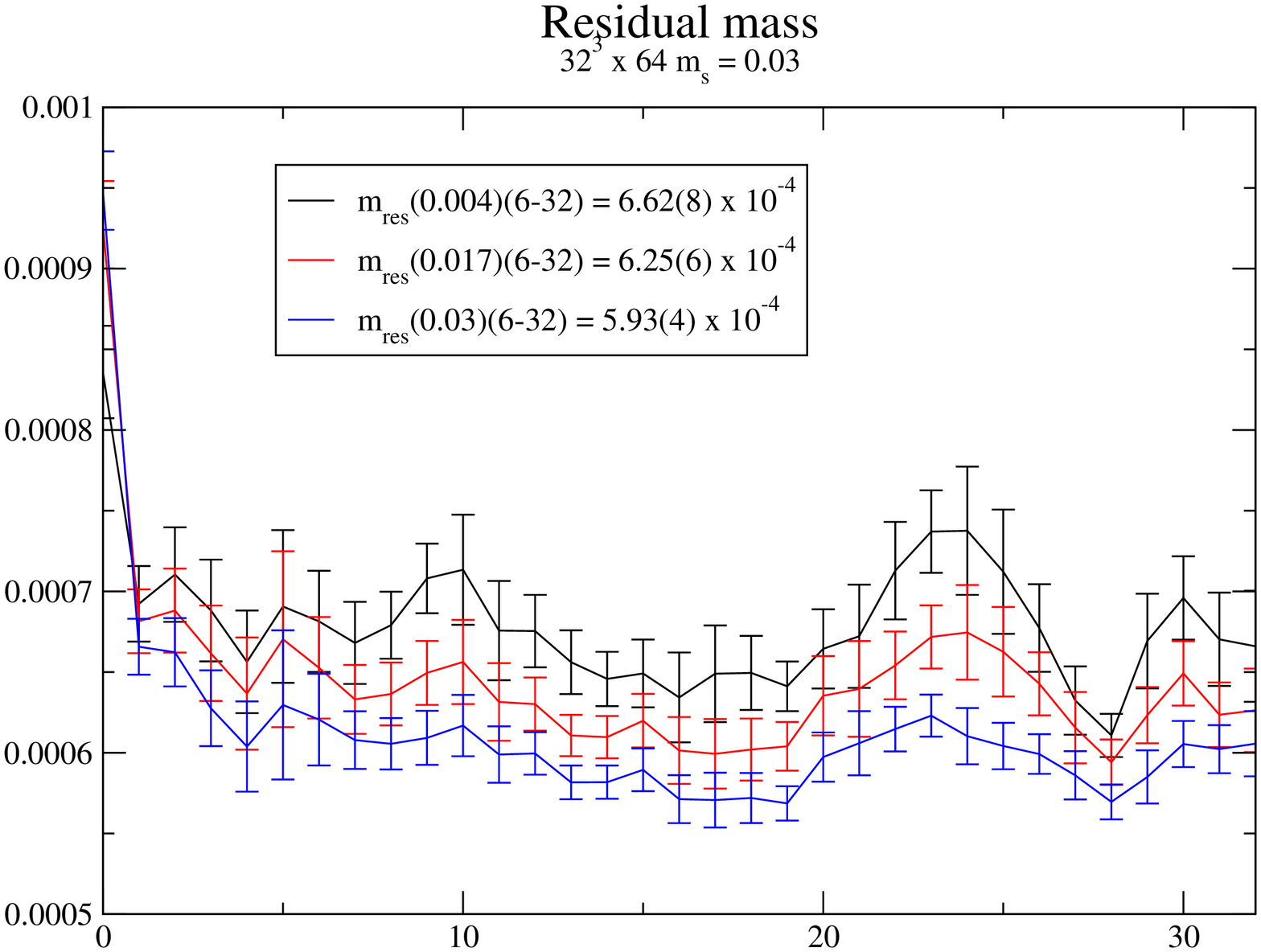}
\end{center}
\end{minipage}
\begin{minipage}{0.5\textwidth}
\begin{center}
\vspace{-5mm}
\includegraphics[angle=270,width=3.3in]{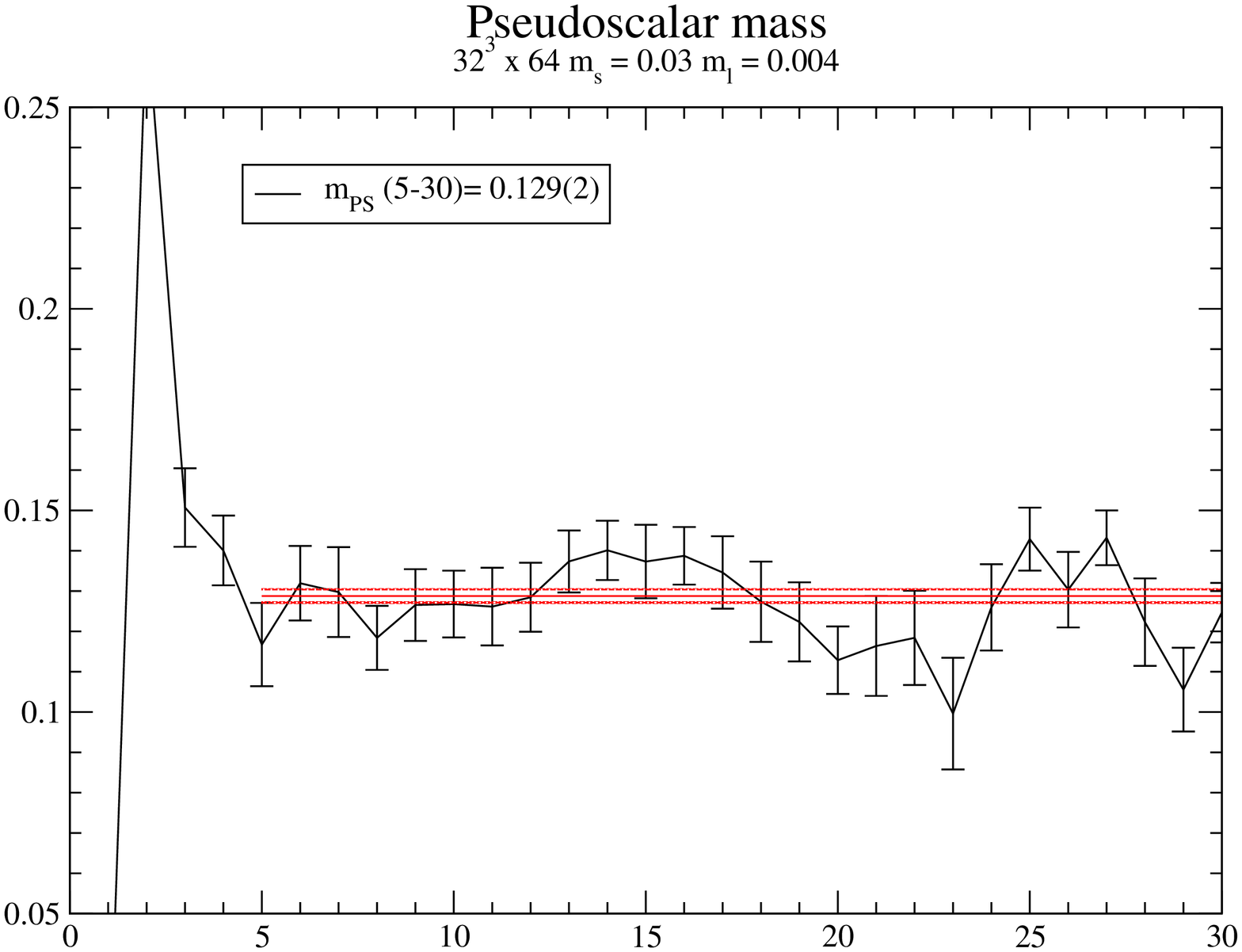}
\end{center}
\end{minipage}
\caption{$R(t)$ for different valence quark masses 
and the pseudoscalar meson effective mass 
on $\beta=2.25, 32^3\times 64 \times 16, m_l=0.004$ ensemble.
Quoted error for $R(t)$ and the mass is from uncorrelated fits, necessary due to long plateaus.}
\label{fig:mres}
\end{figure}

\begin{figure}
\begin{minipage}{0.5\textwidth}
\begin{center}
\includegraphics[angle=270,width=3.3in]{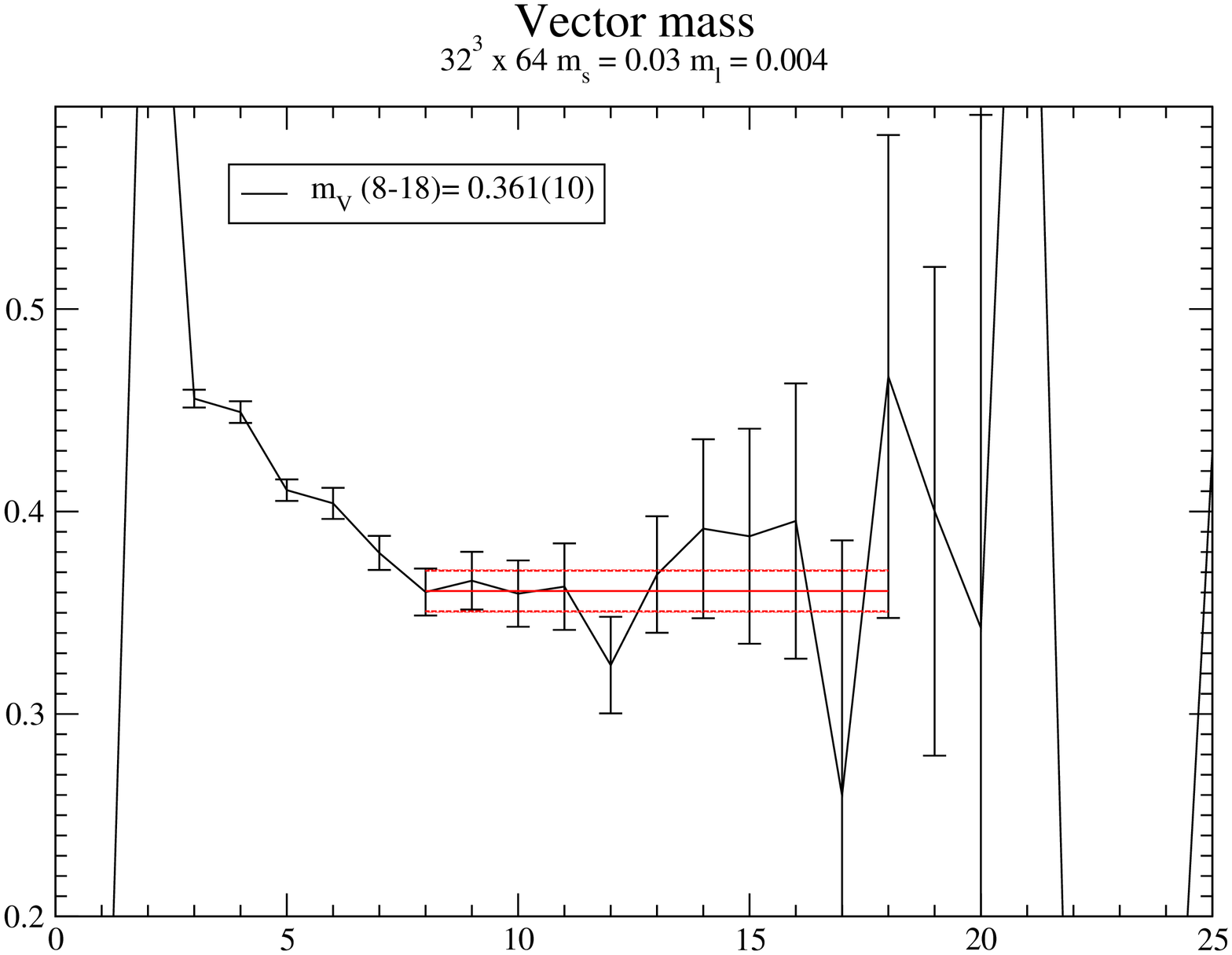}
\end{center}
\end{minipage}
\begin{minipage}{0.5\textwidth}
\begin{center}
\includegraphics[angle=270,width=3.3in]{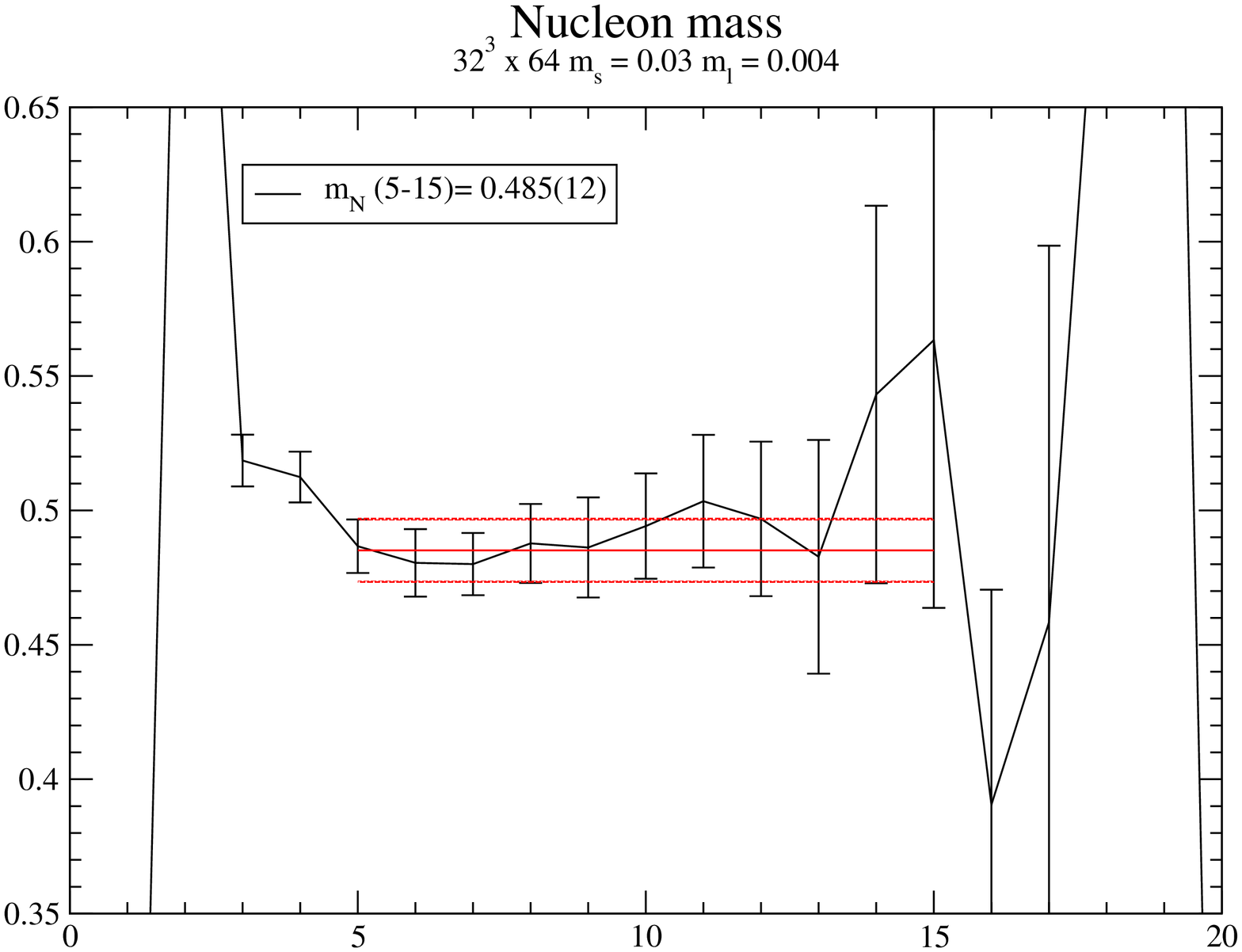}
\end{center}
\end{minipage}
\caption{The vector meson and nucleon effective masses on $m_l=0.004$ ensemble. Error bars are from correlated fits with $\chi^2/d.o.f \sim 1$.}
\label{fig:vec}
\end{figure}

Figures~\ref{fig:mres} and \ref{fig:vec} show the preliminary result 
of the residual masses and various hadron masses measurements. Measurements were done on 30 $m_l=0.004$ lattices from MD trajectory length 300-590 with gauge fixed box sources with size 20, placed at $t=0$ and $t=32$.
This was done mostly to ensure the lattice spacing and residual masses are within estimated range and will be measured again with the sources we will use for other measurements. Chiral extrapolation is not attempted as we measurements on only one dynamical mass.

Residual mass is measured by fitting $R(t)$, a ratio of pseudoscalar and mid-point correlator  defined as
\begin{displaymath} 
R(t)=\frac
{\langle\sum_x J^a_{5q}(x,t)\pi^a(0)\rangle}
{\langle\sum_x J^a_{5}(x,t)\pi^a(0)\rangle}
\end{displaymath} 
to a constant between $t=6$ and 32. 
($J^a_{\{5,5q\}}(x,t)$ are defined in \cite{DWF16Ls8}.)
While the uncorrelated error may be an underestimate of the real error, it shows the residual mass is $\sim 6\times 10^{-4}$ in lattice units  or $\sim$ 1.3Mev. Similarly, fitting meson effective masses gives $a^{-1} \sim$ 2.2Gev. A separate measurement of lattice spacing from the heavy quark potential is in progress.



\section{Conclusions and Discussions}
RBC, UKQCD and LHPC joint collaborations are generating dynamical DWF ensembles with a smaller lattice spacing than which are currently available. These ensembles will reduce the systematic error in continuum extrapolation of many important physics quantities. A preliminary measurements suggests $m_{res} \sim 1/500 m_s, a^{-1} \sim 2.2$Gev and the errors from residual chiral symmetry breaking are expected to be $\sim 10^{-4}$ for $B_k$ and $\sim$ 2 \% for $\epsilon' / \epsilon$, according to the estimate in \cite{sharpe}.

While recent advances in HMC algorithms made gauge configuration generation relatively inexpensive, measurements with multiple valence masses still require significant computational resources.
We are currently working to choose the source which will give optimal overlaps with hadron states we are interested in studying. Also, we are studying various deflation techniques proposed recently.\cite{defl}


\section{Acknowledgments}
The author thanks all the members of the RBC and UKQCD collaborations who contributed to the generation of gauge ensembles and the proceeding.
The computations for this work were performed on the QCDOC machines at University of Edinburgh, Columbia University and Brookhaven National Laboratory. C.J. was supported by the U.S. Dept. of Energy under contract DE-AC02-98CH10886.

\end{document}